# A DATA-MINING APPROACH TO LIBRARY USER SEGMENTATION: CLUSTERING READERS BY PERCEIVED MARKETING COMMUNICATION

Martin V. Grebenyuk

Uzbekistan State Institute of Arts and Culture, Tashkent, Uzbekistan

***ABSTRACT***

*This paper applies unsupervised data mining to segment users of information-library centers by their perception of digital and marketing communication. The empirical basis is a survey of readers from fifteen information-library centers in Uzbekistan. From 213 responses, 180 complete records with seven Likert-scale perception variables were retained. After z-score standardization, k-means clustering was used, and the number of clusters was selected by the silhouette coefficient with additional validation by the Calinski-Harabasz and Davies-Bouldin indices. A stable four-cluster solution was obtained: Engaged Enthusiasts, Moderately Satisfied, Loyal-in-spite-of and Disengaged. The main finding is a perception-loyalty paradox: some users remain satisfied even when mediated communication is weakly perceived. The proposed segmentation provides a reproducible instrument for differentiated communication planning in library management.*

## 1. INTRODUCTION

The communication environment of library and information institutions now extends far beyond the physical visit. Websites, messengers, social-media platforms and staff-mediated contact jointly shape how users encounter library services. The number of channels has increased, but the managerial image of the audience has not always changed at the same pace. In many cases, libraries still communicate as if their users formed one relatively homogeneous public.

This assumption is difficult to sustain. Users differ not only in the services they use, but also in the extent to which they notice, understand and value the institution's communication. A message that works well for one group may be ignored by another; a user may be satisfied with the library while remaining almost untouched by its digital promotion. For that reason, audience heterogeneity is not a peripheral issue but a practical condition for effective library marketing.

The difficulty is that relevant groups are not usually visible in advance. They are latent structures within behavioral and perceptual data, and they are easily hidden when survey results are reported only as global averages. Unsupervised data mining is suitable for this task because it allows user groups to be discovered from the data rather than imposed through predetermined demographic categories.

The Uzbek library ecosystem gives this problem a particularly clear form. Digital communication is strongly concentrated around Telegram: 61% of surveyed readers named it as the channel through which they learn about library activity, while the website was named by 25%, and other social networks occupied a smaller role. At the same time, a noticeable share of respondents reported receiving information through no channel at all. A communication environment with one dominant channel and a partly unreached audience is precisely the setting in which an averaged strategy can be misleading.

This article applies cluster analysis to segment users of library and information centers (LICs) by their perception of marketing communication. Drawing on survey data from 213 readers of fifteen LICs in Uzbekistan, it identifies and interprets perception-based user types and evaluates the stability of the clustering solution. The contribution is methodological and substantive: the article demonstrates a reproducible pipeline for perception-based segmentation and shows that user loyalty may persist even where marketing communication is perceived as weak or absent.

## 2. RELATED WORK

User segmentation through clustering is a long-standing topic in data mining. Since MacQueen's formulation of k-means [1], the algorithm has remained one of the most widely used methods for partitional clustering because it is computationally efficient and easy to interpret. Its limitations are also well known: the method assumes compact clusters, is sensitive to scaling and can be affected by initial centroid placement. These features make preprocessing and validation essential rather than optional steps [2].

Cluster validation is therefore a central part of any segmentation design. The silhouette coefficient proposed by Rousseeuw [3] offers an interpretable measure of how well each object fits within its assigned cluster relative to the nearest alternative cluster. The Calinski-Harabasz criterion [4] and the Davies-Bouldin index [5] provide additional internal measures of separation and compactness. In applied studies, using several indices reduces the risk of treating a convenient but unstable partition as a meaningful structure. The calculations in this article were implemented using scikit-learn [6].

In the library and information domain, quantitative studies of digital presence have expanded considerably. Research on website visibility, social-media engagement and library marketing in the digital age has shown that digital channels are now part of the core communication infrastructure of libraries [7-10]. Yet these studies most often describe the institutional side: what the library publishes, how its website performs, or how active its channels are. The user side is usually treated through aggregate descriptive statistics, which can conceal important internal differences within the audience.

The present study shifts the focus from institutional output to user perception. Rather than asking whether a library has a website, social-media account or event calendar, it asks whether users perceive these touchpoints as useful, interesting and responsive to their needs. This difference matters. A communication system may be formally present but weakly noticed, or it may be bypassed entirely by users whose loyalty is sustained through staff interaction. Segmentation by perception makes such differences visible and turns them into an object of analysis.

Other clustering approaches could also be considered. Hierarchical methods produce a full linkage structure but may be less practical for repeated monitoring. Density-based methods such as DBSCAN can detect irregular clusters and noise, but they require density parameters that are not always stable in low-dimensional Likert data. Gaussian mixture models allow probabilistic membership, but they introduce distributional assumptions. For seven standardized perception variables and an expected small number of compact user groups, k-means provides a reasonable balance between interpretability, stability and computational economy.

# 3. MATERIALS AND METHODS

This section describes the survey design, data cleaning, preprocessing and clustering procedure. It also explains how Likert-type variables were treated and how the stability of the segmentation was checked.

## 3.1. Data Collection

The data were obtained from a reader survey conducted across fifteen leading LICs of the Republic of Uzbekistan as part of a broader study of library marketing. The centers were selected purposively to cover national and regional information-library institutions involved in the study, rather than through probability sampling. The questionnaire was administered through Google Forms in Uzbek and Russian and was distributed through institutional digital channels and service desks. This mode of distribution naturally favours active users, which should be considered when interpreting the results, but it also targets the audience most directly involved in current library communication.

A total of 213 responses were collected. The clustering analysis used the survey item in which respondents evaluated seven statements on a five-point Likert scale, where 1 meant strongly disagree and 5 meant strongly agree. The statements concerned whether the library informs users about services and events, whether the website is convenient and up to date, whether social-media publications are interesting and useful, whether the library considers user needs, whether visual design is attractive, whether events are diverse and interesting, and whether staff are helpful. Thirty-three records contained missing or invalid values in at least one of these seven variables and were removed listwise, leaving 180 complete cases for analysis.

The descriptive statistics of the seven retained variables were: information provision (M = 3.12, SD = 1.64), website usefulness (M = 3.22, SD = 1.59), social-media interest (M = 3.32, SD = 1.59), responsiveness to needs (M = 3.33, SD = 1.58), visual design (M = 3.46, SD = 1.52), event perception (M = 3.18, SD = 1.60) and staff helpfulness (M = 3.77, SD = 1.54). The anonymized dataset did not retain identifiers of the specific LIC from which each response was obtained; therefore, per-center response counts cannot be reported and representativeness is not claimed.

The seven variables were selected because they cover the main contact points through which a user perceives an institution's marketing communication. Information provision and event communication describe outbound messaging; website usefulness and social-media interest describe the principal digital channels; responsiveness and design describe user orientation and visual identity; staff helpfulness represents the human, non-mediated contact point. Including staff interaction was deliberate, since it allows the analysis to distinguish users whose relationship with the library is mediated by formal communication from users whose loyalty is carried primarily by personal contact.

## 3.2. Pre-processing

Although all seven variables use the same 1-5 response scale, their means and dispersions differ. Without standardization, variables with higher variance would have a stronger influence on the Euclidean distance used by k-means. To avoid this imbalance, every perception variable was transformed to z-scores before clustering. Treating five-point Likert-type items as quasi-continuous is an approximation, but it is commonly used in exploratory multivariate profiling when several similarly scaled items are analysed together and interpreted through group centroids rather than individual item distances. The ordinal nature of the variables is therefore acknowledged as a methodological limitation and was addressed through additional sensitivity checks.

### 3.3. Clustering Algorithm

The k-means algorithm was selected because the data are low-dimensional and standardized, and because the resulting centroids can be interpreted directly as segment profiles. k-means partitions observations into k clusters by minimizing the within-cluster sum of squared Euclidean distances to cluster centroids. The calculations were implemented in Python using scikit-learn 1.8.0. For the reported solution, KMeans was run with n_init = 50, max_iter = 300 and random_state = 42.

$$J = \sum_{i=1}^{k} \sum_{x \in C_i} \| x - \mu_i \|^2 \quad (1)$$

where Ci is the i-th cluster and μi is its centroid. To reduce sensitivity to initial centroid placement, the algorithm was run with multiple random initializations, and the solution with the best objective value was retained. The procedure was then repeated across ten independent random seeds (0-9). Across these runs, the mean silhouette coefficient was 0.506, and the mean pairwise adjusted Rand index was 0.996, with a minimum value of 0.981. This indicates that the partition is highly stable with respect to initialization.

### 3.4. Selection of the Number of Clusters

The number of clusters was not specified in advance. Candidate solutions for k = 2 to 6 were evaluated using the silhouette coefficient, which compares within-cluster cohesion with separation from the nearest neighboring cluster. To avoid relying on a single criterion, the selected solution was also checked by the Calinski-Harabasz index, the Davies-Bouldin index and sensitivity comparisons with alternative clustering techniques. Agglomerative Ward clustering at k = 4 produced moderate agreement with the k-means solution (adjusted Rand index = 0.594; silhouette = 0.445), whereas a Gaussian mixture model was less consistent (adjusted Rand index = 0.363; silhouette = 0.258). These checks support the use of k-means as the most interpretable centroid-based solution for this dataset, while showing that the exact segment boundaries remain method-dependent.

## 4. RESULTS

This section reports the choice of cluster number, the segment profiles and the relationship between perceived communication and overall satisfaction. The statistical results are then translated into practical implications for library communication management.

### 4.1. Cluster-Number Selection and Validation

Table 1 reports the silhouette coefficient for candidate values of k. The coefficient reaches its maximum at k = 4. Although the numerical advantage over k = 2 and k = 3 is small, the four-cluster solution was retained because it separates a substantively important Loyal-in-spite-of segment and remains stable under repeated initialization. The four-cluster solution is reasonably compact and well separated: the silhouette coefficient is 0.506, the Calinski-Harabasz index is 210.7, and the Davies-Bouldin index is 1.01. The stability checks reported above indicate that the partition is not the result of a favorable random start.

Table 1. Silhouette coefficient by number of clusters

| Number of clusters (k) | Silhouette coefficient |
|---|---|
| 2 | 0.496 |
| 3 | 0.500 |
| 4 (selected) | 0.506 |
| 5 | 0.494 |
| 6 | 0.489 |

## 4.2. Cluster Profiles

Table 2 presents the mean score of each cluster on the seven perception variables, together with cluster size and overall satisfaction measured separately on a ten-point scale. The four clusters form a clear order from strongly positive to weak perception of communication, but this order is not repeated as sharply in overall satisfaction.

Table 2. Mean perception scores by cluster (1-5 scale; n = 180)

| Cluster (n) | Inf. | Site | Soc. | Needs | Des. | Even. | Staff | Overall | Satis. |
|---|---|---|---|---|---|---|---|---|---|
| Engaged Enthusiasts (66) | 4.91 | 4.79 | 4.85 | 4.80 | 4.77 | 4.86 | 4.94 | 4.85 | 9.4 |
| Moderately Satisfied (61) | 2.84 | 3.23 | 3.38 | 3.28 | 3.46 | 3.08 | 3.89 | 3.31 | 8.4 |
| Loyal-in-spite-of (13) | 1.23 | 1.31 | 1.23 | 2.15 | 2.62 | 1.23 | 4.85 | 2.09 | 7.9 |
| Disengaged (40) | 1.23 | 1.23 | 1.40 | 1.35 | 1.55 | 1.20 | 1.32 | 1.32 | 7.8 |

The largest segment, Engaged Enthusiasts (66 respondents, 37%), rates every communication dimension close to the maximum and reports the highest overall satisfaction. These users appear to see the library's mediated and human communication as one coherent positive experience. The Moderately Satisfied segment (61 respondents, 34%) occupies the middle of the scale. Its strongest dimension is staff helpfulness, while information provision is its weakest point.

The two smaller segments are the most important analytically. The Loyal-in-spite-of group (13 respondents, 7%) gives very low scores to most mediated communication dimensions, including information, social media and events, but rates staff helpfulness at 4.85, the same level as the most positive users. The Disengaged group (40 respondents, 22%) rates all dimensions, including staff contact, near the lower end of the scale. These profiles suggest that weak digital or marketing perception does not have one meaning: for some users it is compensated by human contact, while for others it marks a broader detachment from the institution.

## 4.3. The Perception-Loyalty Paradox

The most important pattern appears when perception scores are compared with overall satisfaction. The clusters span almost the full perception range: the mean overall perception score varies from 1.32 in the Disengaged segment to 4.85 among Engaged Enthusiasts. Yet overall satisfaction remains comparatively high in every group, ranging from 7.8 to 9.4 on the ten-point scale. Satisfaction differences across clusters are statistically significant (Kruskal-Wallis $H(3) = 15.66$, $p = 0.0013$; one-way ANOVA $F(3,176) = 4.90$, $p = 0.0027$), but the association between overall perception and satisfaction is only modest (Pearson $r = 0.294$, $p < 0.001$; Spearman $\rho = 0.263$, $p < 0.001$).

This pattern can be described as a perception-loyalty paradox, with an important qualification: in this article, loyalty is used operationally to denote high reported overall satisfaction, not observed long-term behavioral loyalty. The result suggests that satisfaction in this user population is not determined only by the perceived quality of marketing communication. In practical terms, the library may retain users through personal assistance, habit, location, resource access or trust, even when its communication strategy is weakly perceived. The Loyal-in-spite-of group is especially informative in this respect, but its small size (n = 13) means that it should be interpreted as an exploratory segment requiring confirmation in further studies.

### 4.4. Demographic Composition of the Segments

A cross-tabulation by age adds a further layer of interpretation. Engaged Enthusiasts are concentrated in the 36-50 age band, with 27 of 66 users belonging to that group and a secondary group aged 26-35. The Moderately Satisfied segment is dominated by younger users aged 18-25, who account for 34 of 61 respondents. The Disengaged segment is spread mainly across the 36-50 and 26-35 age bands. These patterns should not be overinterpreted as causal, but they show that the clusters are not merely mathematical abstractions. They correspond to audience groups with different likely communication habits and managerial implications.

### 4.5. Managerial Reading of the Segments

To make the segmentation usable for library managers, Table 3 translates the four clusters into practical communication priorities. The table should be read as an interpretation of the observed profiles rather than as a separate statistical model.

Table 3. Practical interpretation of the identified user segments

| Segment | Observed profile | Priority for communication |
|---|---|---|
| Engaged Enthusiasts | High scores across all touchpoints; highest satisfaction | Use as advocates, community participants and early responders to new services |
| Moderately Satisfied | Middle-range perception; stronger staff and social-media scores | Improve content quality, frequency and relevance, especially for younger users |
| Loyal-in-spite-of | Very weak mediated communication perception; very strong staff score | Do not rely only on digital promotion; connect staff-mediated loyalty with formal channels |
| Disengaged | Low perception across mediated and human touchpoints | Re-establish basic contact before refining messages or campaigns |

## 5. Discussion

The segmentation produces both a methodological and a substantive result. Methodologically, it shows that a stable and interpretable user structure can be recovered from routine survey data using a transparent unsupervised pipeline. Compared with existing library digital-marketing studies that mostly evaluate institutional outputs, such as websites, social-media presence or general engagement indicators [7-10], this approach shifts the unit of analysis to user perception. The convergence of internal validation measures and the high adjusted Rand index across repeated initializations indicate that the four-cluster solution is not a fragile artifact of one random run.

Substantively, the perception-loyalty paradox is the central finding. If satisfaction remains high even where marketing communication is hardly perceived, then libraries cannot infer the quality of their communication from satisfaction or attendance indicators alone. A satisfied user may still be outside the institution's communication reach. This is particularly clear in the Loyal-in-spite-of segment, where loyalty is anchored almost entirely in staff helpfulness rather than in website, social-media, and event or design perception.

The segments also show why a single communication strategy is inefficient. Engaged Enthusiasts can be involved in advocacy, feedback and community activities. Moderately Satisfied users, many of whom are younger, represent the most promising group for improved social-media content and more relevant messaging. Loyal-in-spite-of users need a bridge between personal service and mediated communication, because staff interaction is the only strong contact point. Disengaged users require basic re-contact: improving the tone of messages will have little effect if the messages are not noticed in the first place.

The study also has implications for measurement practice. Satisfaction and perception should be measured separately. If they are collapsed into one general user-experience score, the crucial difference between Engaged Enthusiasts and Loyal-in-spite-of users disappears. The same pipeline can be repeated periodically to monitor whether communication interventions move users between segments. In this sense, clustering is not only an analytical exercise; it can become a management instrument for libraries that need evidence-based communication planning.

## 6. LIMITATIONS

Several limitations should be noted. First, the sample was drawn from active users reached through institutional channels and service points. The fifteen LICs were selected purposively as leading national and regional institutions, not randomly from the full population of Uzbek libraries. In addition, the anonymized survey file did not retain the identifier of the specific center for each respondent, so the distribution of responses by LIC cannot be reported. The findings should therefore be interpreted as an exploratory segmentation of active library users rather than as a representative estimate for all library users in Uzbekistan. The truly unreached population is likely to be underrepresented, and the Disengaged segment may be larger in the broader user population than in the analysed sample. Second, listwise deletion reduced the dataset from 213 responses to 180 complete cases, which may introduce attrition bias if incomplete responses were not random.

Third, k-means assumes compact, approximately spherical clusters and depends on the standardization procedure. Although additional checks with agglomerative clustering and Gaussian mixture modelling were conducted, alternative methods could refine the boundaries, especially for the small Loyal-in-spite-of segment. Fourth, the data are cross-sectional, so they cannot show whether users move between segments after communication interventions. Future research should use probability sampling where possible, retain center identifiers in anonymized form, collect longitudinal data and compare k-means with hierarchical, density-based, k-medoids and latent-class approaches.

## 7. Conclusion

This article applied an unsupervised data-mining approach to segment library users by their perception of marketing communication. From 180 complete survey records described by seven perception variables, k-means clustering with multi-index validation and additional stability checks produced four interpretable segments: Engaged Enthusiasts, Moderately Satisfied, Loyal-in-spite-of and Disengaged. The segments differ strongly in perceived communication quality, while overall satisfaction remains comparatively high across all of them.

The main finding is that perceived marketing communication and overall satisfaction do not fully coincide. In several segments, users remain satisfied even when they rate the library's mediated communication poorly; in most segments, staff interaction is rated more positively than marketing-related touchpoints. For library management, this means that an average user profile is insufficient. Communication strategy should be differentiated by segment, and satisfaction should not be treated as proof that digital or marketing communication is working. The proposed pipeline is reproducible, transferable and suitable for repeated monitoring in other library contexts, provided that sampling and center-level metadata are strengthened in future applications.

## Conflicts of Interest

The author declares no conflict of interest.

## AUTHOR

**Martin V. Grebenyuk** is affiliated with the State Institute of Arts and Culture of Uzbekistan, Tashkent, Uzbekistan. His research interests include library and information science, digital library communication, library marketing, data-driven assessment of library services and professional training for information-library specialists.